\documentclass[12pt,preprint]{aastex}

\slugcomment{To Appear in ApJ, 2008}

\shorttitle{Phase-Resolved Cyclotron Spectroscopy of Polars}
\shortauthors{Campbell et al.}

\begin{document}

\title{Cyclotron Modeling Phase-Resolved Infrared Spectroscopy of Polars III: 
AM Herculis and ST Leo Minoris}

\author{Ryan K. Campbell\altaffilmark{1,2} and
Thomas E. Harrison\altaffilmark{1,3}}
\affil{Astronomy Department, New Mexico State University,
  Las Cruces, NM 88003}

\author{Stella Kafka}
\affil{Spitzer Science Center / Caltech, MS 220-6, 1200 E.California Blvd, 
Pasadena, CA 91125}

\altaffiltext{1}{Visiting Astronomer, Infrared Telescope Facility.
IRTF is operated by the National Aeronautic and Space Administration}
\altaffiltext{2}{Supported by New Mexico State Space Grant}
\altaffiltext{3}{Visiting Astronomer, Kitt Peak National Observatory}

\begin{abstract}
We present phase-resolved low resolution infrared spectra of
AM Her and ST LMi, two low-field polars that we observed with SPEX on the IRTF.
Optical/NIR lightcurves are also published to help constrain the viewing 
geometry and brightness of the objects at the time they were observed.
Currently, only limited IR spectra have been published for these
objects, and none with the phase-coverage presented here. 
In both cases, the resulting spectra are dominated by emission from the 
secondary star in the NIR. However, the emission regions are 
also self-eclipsed, allowing us to isolate the cyclotron emission through
subtraction of the dim-phase spectrum. We use a ``Constant Lambda''
prescription to model the changing cyclotron features seen in the resulting
data. For AM Her, we find a best fit model of: B = 13.6 MG, kT = 4.0 keV,
and log$\Lambda$ = 5.0. The cyclotron derived accretion geometry is 
consistent with $i$ = 50$^{\circ}$ and $\beta$ = 85$^{\circ}$. 
For ST LMi, B = 12.1 MG, kT = 3.3 keV, and log$\Lambda$ = 5.7 
with $i$ = 55$^{\circ}$ and $\beta$ = 128$^{\circ}$.

\end{abstract}


\section{Introduction}

Polars are interacting binary systems containing a primary white dwarf (WD) and
a late type secondary star. Material flows from the 
secondary, through the L$_{1}$ point and falls ballistically toward the WD. 
The WDs in polars are highly magnetized with magnetic field strengths that 
range between 10 - 240 MG. Eventually, the accreting material couples to 
the field lines of the WD and is transported to the magnetic pole(s) of the 
star where a dense, standing-shock is formed, with nominal temperatures of 
2 - 20 keV, which cools by emitting bremsstrahlung and cyclotron radiation . 

AM Her is the prototype polar (see Tapia, 1977a). It has an
orbital period of 3.094 hrs and is nearby (78 pc; Thorensten et al., 2003). 
Despite being extensively 
studied many characteristics of the system remain uncertain.
One point of discrepancy is the exact mass of the primary WD with
estimates ranging from 0.39 M$_{\odot}$ (Young et al., 1981) to 1.22 
M$_{\odot}$ (Cropper et al., 1998). Additionally, the geometry of the 
system remains unclear with orbital inclination 
estimates ranging from $i$ = 35$^{\circ}$ (Brainerd \& Lamb, 1985) to $i$ = 
60$^{\circ}$ - 80$^{\circ}$ (Watson et al., 2003), although the self-eclipse 
observed in the X-Ray and UV requires that $i$ + $\beta$ $\geq$ 90$^{\circ}$,
where $\beta$ is the magnetic co-latitude. Indeed, both Sirk $\&$ Howell (1998)
and Gansicke et al. (1998) found that  
$i$ + $\beta$ = 105$^{\circ}$. Later, Gansicke et al.(2001) determined that 
combinations of $i$ and $\beta$ between ($i$ = 50$^{\circ}$, $\beta$ = 
55$^{\circ}$) and ($i$ = 35$^{\circ}$, $\beta$ = 70$^{\circ}$) best modeled the
high-state optical lightcurves. In other ways, however, AM Her is well 
characterized. The temperature of the primary WD has been well 
constrained. Gansicke et al. (2006) modeled low-state FUSE and STIS spectra 
from AM Her with TLUSTY/SYNSPEC (Hubeny \& Lanz, 1995), finding that 
T$_{wd}$ = 19800 $\pm$ 700 K. Also, the secondary star is
spectroscopically determined to be M4 $\pm$ 1 (Kafka et al., 2005b), although 
there is some evidence that it 
is irradiated by the WD and thus the spectral type of the secondary 
is orbitally modulated (Davey \& Smith, 1992). The published photometry of AM 
Her is exhaustive. Orbital lightcurves show variability in every pass-band 
from the UV (Gansicke et al. 1998) out to the $K$-band (see below). Extensive
AAVSO and automatic photometric telescope, ``APT'' (e.g. RoboScope) monitoring 
from 1990 - 2004, has revealed that AM Her is usually
in one of two states: a ``high-state'', with large intrinsic variability: 13.0 
$\leq V \leq$ 14.0, and a ``low-state'' with $V$ $\simeq$ 15.5 (Kafka et al., 
2005a). Finally, Bailey, Ferrario, \& Wickramasinghe (1991; henceforth BFW91) 
used a Constant Lambda (``CL'') code to model the NIR low-state (V $\sim$ 15.0)
cyclotron spectrum, which were binned into bright-phase ($\phi$ = 0.46 - 0.88) 
and dim-phase ($\phi$ = 0.46 - 0.88) spectra. The dim-phase showed only 
emission from the secondary star, while strong cyclotron emission was observed 
for the duration of the bright phase. After subtracting a 3250 K model 
atmosphere with log $g$ = 4.75 to mimic the secondary spectrum, BFW91 found 
B $\simeq$ 14.5 MG, and a shock temperature of kT = 8.5 keV. 

ST LMi (= CW1103 +254) is a short period polar (P$_{orb}$ = 114 min)
containing a 0.7 M$_{\odot}$ primary (Ramsay et al. 2004) with a
likely temperature of 11000 K (Araujo-Betancor et al., 2005; Sion, 1999)
and a M5 - M6 secondary (Knigge, 2006; Harrison et al., 2005; Howell 
et al., 2000; Warner, 1995) at a distance of 115 - 138 pc (Araujo-Betancor
et al., 2005; Kafka et al., 2007). It was classified as an AM Her object by 
Stockman et al. (1983) on the basis of its highly variable polarization. For 70
\% of the orbit the object shows no significant polarization. 
Subsequently, a strong pulse is observed peaking near $\phi$ = 0.00 at 12 \% 
and -20 \% in linear and circular light, respectively (Cropper et al., 1986). 
This observed bi-modality is echoed in optical and IR orbital lightcurves, 
which show a quiescent ``dim-phase'' for most of the orbit in each band that 
is followed by a significant jump in brightness (the ``bright-phase'') 
coincident to the peak in polarization. Peacock et al. (1992) obtained 
multi-band photometry of ST LMi, showing that $\Delta$m $\simeq$  0.6, 1.4, 
2.0, 1.5, and 1.3 for the $BRIJH$ bands, respectively. Long-term $V$-band
lightcurves were obtained with RoboScope from 1990 to 2003 (Kafka et al., 
2005a). From 1992 - 1997, the system was in a protracted ``low-state'' with  
$<$$V$$>$ = 17.5 $\pm$ 0.2. From 1997 - 2003, a more variable, slightly higher 
state was observed with $<$$V$$>$ = 16.0 $\pm$ 1.5. Additionally, instances of
``extreme low-states'' have been observed. In Ciardi et al. (1998)
the $K$-band spectra of ST LMi, showed no obvious emission lines and were 
modeled successfully with a $\simeq$ 3000 K atmosphere 
suggesting that accretion had almost completely shut off. Kafka et al. (2007) 
has presented photometry of a similar extreme low-state showing that the 
system can be as faint as $V$ = 18.5. $JHK$ cyclotron spectra were previously 
modeled by Ferrario, Bailey \& Wickramasinghe (1993; henceforth FBW93) finding 
that two spots were necessary to fully model their spectra: a primary region 
with B = 12.0 and kT = 12 keV, and a secondary, ``cool-spot'' with  kT = 5.0 
keV. The accretion geometry of the primary emission region has also been
previously determined, with 55$^{\circ}$ $\leq i \leq$ 64$^{\circ}$ and 
140$^{\circ}$ $\leq \beta \leq$ 150$^{\circ}$ (Schmidt et al., 1983; 
Potter, 2000). 

Limited IR spectroscopy exists for these two low-field polars.  Below, we 
present and model new phase-resolved low-state infrared spectra as well 
as $JHK$ lightcurves for AM Her and ST LMi. In both instances, 
we show that variable cyclotron emission over the orbit is responsible for 
the spectroscopic and photometric behavior. Additionally, we present a second 
epoch data set for ST LMi that shows no cyclotron emission and must be in
an extreme low-state similar to that seen by Kafka et al. (2007). In the next 
section, 
we describe the observations of each object, in section 3 we fit these data 
with cyclotron models, discuss our results in section 4, and draw our 
conclusions in section 5.

\section {Observations}

AM Her and ST LMi were observed using SPEX (c.f. Rayner et al., 2003) on the 
Infrared Telescope Facility (IRTF). AM Her was observed once on 2005 September 
1, whereas ST LMi was observed on two different epochs: 2005 Feb 7 and 2006 
Feb 2. Both AM Her and ST LMi were found to be in low-states, although as we 
discuss below, the 2005 Feb 7 data found ST LMi in an extreme 
low-state. SPEX was used in low-resolution ``prism'' mode with a 0.3'' x 15'' 
slit. To remove background, each object was nodded along the slit. In its 
low-resolution mode SPEX produces R(=$\lambda$/$\Delta \lambda$) 
$\sim$ 250 spectra, with short enough exposure times to obtain phase resolved 
spectra of polars with $K \leq 16.0$.  For ST LMi, we used 240 second exposure 
times, where shorter, 120 s, integration times were adequate for AM Her. Each 
of these spectra were then median combined with 2-3 other spectra to allow for 
cosmic-ray removal and to improve the S/N ratio. The spectra were reduced 
using the SPEXTOOL package (Vacca et al., 2003). A telluric correction was 
applied using an A0V star of similar airmass to our program objects. We use 
the Kafka et al. (2005b) ephemeris to phase all observations of AM Her. For ST 
LMi because of the large phase uncertainty ($\Delta \phi \simeq$ 0.10) in the 
Howell et al. (2000) ephemeris, we phased our observations to the $J$-band 
minimum found in the photometry presented in this paper, which worked out to a 
phase-shift of $\Delta \phi$ = 0.15 from that ephemeris.

Because of the narrow slit size on IRTF/SPEX (0.3'') infrared photometry
is required to calibrate the fluxes of the spectra. The $JHK$ photometry for 
each object was obtained with SQIID on the KPNO 2.1-m telescope (Ellis et al., 
1992). AM Her was observed on 2002 September 26, and ST LMi on 2003 April 9. In
addition, we obtained simultaneous $BVRIJHK$ photometry on 2005 May 20 for 
AM Her. The $JHK$ photometry was obtained with NIC-FPS\footnote{see 
http://www.apo.nmsu.edu/arc35m/Instruments/NICFPS/nicfpsusersguide.html} 
on the Apache Point 3.5-m, while the optical data set was obtained with the 
NMSU 1-m (see Harrison et al., 2003). To aid the reader, we have collated
all the observational specifics in Table 1.

\section{Modeling}

To produce our cyclotron models, we use a Constant-Lambda (``CL'') 
cyclotron code first developed by Schwope (1990). In Campbell et al. (2008a; 
hereafter paper I), we presented a theoretical synopsis of CL modeling which 
will not be repeated here. The
model spectra depend on four global parameters: B (the magnetic field 
strength), kT (the plasma temperature), $\Theta$ (the viewing angle to the 
magnetic ``pole''), and $\Lambda$ (the ``size parameter''), which is 
closely tied to the column density along the line of sight through the 
accretion region. In paper I, we found that we could adequately model the
data for EF Eri as cyclotron + WD. In Campbell et al. (2008b; hereafter paper 
II), we found that in many polars there are other sources of non-stellar
continuum radiation (e.g., Bremsstrahlung emission) which contaminate the 
spectra and need to be taken into account. For each object in paper II, 
the accretion column was self-eclipsed. In this case, cyclotron emission is 
only seen for the part of the orbital cycle when the accretion column is in 
view (the ``bright-phase'') although it is contaminated by other sources. To 
subtract these away, we assume that the ``dim-phase'' spectra represent all 
the additional components of radiation which obfuscate the cyclotron emission. 
The dim-phase spectrum is then subtracted from the spectra at other phases 
where accretion column is in view, thus yielding uncontaminated cyclotron 
spectra over the orbit. We refer to the dim-phase subtraction method as 
``Stream-Emission Subtraction'' (SE-subtraction) for consistency with Schwope 
et al. (2002). An additional contaminant for the objects in the current work 
is the irradiated secondary star whose spectral type slowly changes with phase
and can not be completely subtracted. Thus, when the SE-subtraction technique 
was applied features due to the secondary star remained.

\subsection{AM Her} 

In Figure 1, we show the $JHK$ lightcurves taken with the SQIID on the KPNO 
2.1-m on 2002 September 26 , at a time when the system was at $V$ = 15.5. This 
is a typical low-state magnitude identical to that of our SPEX spectroscopy, as
shown in Fig. 2. Overlaid in each band, are binary star models computed 
using WD2005\footnote{WD2005 is an updated version of WD98, and can be obtained
at this website maintained by J. Kallrath: 
http://josef-kallrath.orlando.co.nz/HOMEPAGE/wd2002.htm} with a M5 secondary
at an orbital inclination of $i$ = 50$^{\circ}$. The $J$-band morphology is 
well explained by classic ellipsoidal variations except the lightcurve minimum 
at $\phi$ = 0.00 is somewhat deeper than predicted and residual structure 
appears at the $\Delta$J = 0.05 mag level. The derived inclination
should be considered a lower limit because other dilution sources may be 
present. Both the $H$ + $K$ lightcurves show a large cyclotron component 
folded-in with the ellipsoidal variations. In Fig. 3, we include additional 
$BVRIJHK$ photometry obtained 2.5 years later on 2005 May 20, at a time when 
the system was again at similar brightness. We find that the same ellipsoidal 
models provide excellent fits to the $JHK$ data at this epoch. Because 
of the narrow slit size on SPEX (0.3''), at each orbital phase we flux 
calibrate our spectra to the ``cyclotron-free'' $J$-band lightcurve.

The IRTF phase-resolved spectra from 2005 September 1 are dominated by 
emission from the secondary star. To remove this component, we
subtracted the spectrum at $\phi$ = 0.42 from every other phase. The 
subtraction spectrum is near to both the ellipsoidal minimum ($\phi$ = 0.50) 
and because of the ongoing self-eclipse, is free of cyclotron emission. 
To approximate the effect of the ellipsoidal variability,
we scaled the subtraction spectrum at each phase to match the magnitude 
expected from our ellipsoidal models. The underlying continuum SED and 
intrinsic water vapor features at 1.35 and 1.85 $\mu$m in the residual 
spectra were orbitally variable producing a small blue excess and apparent 
water vapor emission at ellipsoidal maxima ($\phi$ = 0.25, 0.75), and a red 
excess with apparent water vapor absorption at ellipsoidal minima ($\phi$ = 
0.00, 0.50), resulting from a changing spectral type as the distorted secondary
star changed orientation. From M4 to M6 water vapor absorption becomes ever 
more pronounced. Thus, even small differences of the secondary temperature are 
apparent in our data and residuals from the water vapor features remain in our 
final spectra (shown in black in Fig. 4). The final cyclotron models are 
overlaid in green. No cyclotron emission was observed over the 
interval 0.27 $\leq \phi \leq$ 0.74 due to the self-eclipse of the emission 
region. Thus, these phases are not shown in Fig. 4 to aid in the presentation 
of data. Between 0.92 $\leq \phi \leq$ 0.09, the spectra appear to show a 
single strong cyclotron harmonic ($n$ = 4) near 2.0 $\mu$m. The $n$ = 4, 5, and
6 harmonics are obvious between 0.20 $\leq \phi \leq $ 0.26 and again from 
0.75 $\leq \phi \leq $ 0.86.

For our best fit cyclotron models at each phase, see Table 2. The average 
parameter values are: B = 13.6$_{-0.8}^{+1.0}$ MG, kT = 4.0$_{-1.0}^{+1.5}$ 
keV, and log$\Lambda$ = 5.0$_{-0.6}^{+0.6}$, with an average $\chi^{2}_{\nu}$ 
= 2.42. The statistical limits were derived by finding where the value of 
$\chi^{2}_{\nu}$ changed by 50 \% over its fiducial value. We find that 
$i$ + $\beta$ = 135$^{\circ}$, with $i$ = 50$^{\circ}$, $\beta$ = 85$^{\circ}$,
$\phi_{min}$ = 0.01, where $\phi_{min}$ is defined as the bluest position of 
the cyclotron harmonics.

\subsection{ST LMi}

In Fig. 5, we present $JHK$ lightcurves of ST LMi taken with the KPNO 2.1-m 
during the normal high accretion state of the system. 
In each band the morphology is similar: the dim-phase lasts from 0.00 
$\leq \phi \leq$ 0.55 with mean magnitudes of 14.4, 13.9, and 13.9 in 
the $J$, $H$, and $K$-bands, respectively. During the subsequent bright-phase, 
the object brightens significantly ($\Delta J$ = 1.5 mag). Like AM Her, a 
WD2005 ellipsoidal model was fit to the NIR lightcurves, finding 
$i$ = 55$^{\circ}$. The models well approximate the dim-phase of ST LMi 
in the $J$ and $H$-bands, while the fit in the $K$-band is more uncertain due 
to the larger scatter in the photometry.

ST LMi was observed spectroscopically over its entire orbit once on 2005 
February 7 and once on 2006 February 2. On both occasions, photometry is 
available within one month of our phase-resolved spectroscopy. Fig. 6 shows 
the high/low states of ST LMi over the entire history of RoboScope and also 
over a shorter, $\sim$ 1.5 year baseline surrounding our 2005 and 2006 IRTF 
observations. For the first dataset, the nearest RoboScope data (February 28) 
shows 17.1 $\leq V \leq$ 17.8, similar in brightness to its 1992 - 1997 
protracted ``low-state''. However, as will be discussed below, our bright-phase
spectroscopy show a conspicuous lack of cyclotron emission during this epoch.
For this reason, we believe the object was in fact in an ``extreme low-state'' 
at the time of observation, similar to that observed on 2006 Feb 12 by Kafka 
et al. (2007) which found 18.0 $\leq V \leq$ 18.4. Curiously, our 2006 data 
was obtained only 10 days prior to that epoch, but shows clear 
evidence of cyclotron emission and thus must have been in a normal low-state.
We note, however, photometric extreme low-states like that observed by Kafka et
al. are short lived, as normal low states were observed within a month both 
before and after it.
  
The spectra of ST LMi are also strongly contaminated by its secondary star at 
every phase. During the extreme low-state we found no evidence for cyclotron 
emission, with the secondary contributing all of the NIR flux at each 
orbital phase. In Fig. 7, we show the observed SPEX spectrum at $\phi$ = 0.02 
of the 2005 dataset with the best-fitting secondary template overlaid (M6).

For the 2006 low-state, we present SE-Subtracted data in Fig. 8a. The SE 
spectrum was produced by averaging the dim-phase spectra together. Since the 
relative uncertainty in the Howell et al. (2000) ephemeris is rather large 
($\Delta\phi_{0} \simeq 0.10$), we phased our data
by defining ellipsoidal minimum in our 2003 KPNO lightcurve as $\phi$ = 0.50,
and then using the Howell et al. (2000) period. We found that averaging three
dim-phase spectra together produced the best SE-subtraction spectra, with 
faint cyclotron features visible during the bright orbital phase at $\sim$ 
2.25, 1.85, 1.53 and perhaps 1.30 $\mu$m, corresponding to the $n$ = 4 - 7 
harmonics in a field with B = 12 MG. Like AM Her, the tidally distorted nature 
of the secondary star imparts spectral type and overall flux changes that are 
orbitally modulated allowing residual emission/absorption from the secondary 
star to remain even after SE subtraction has been performed. Indeed, the upturn
in the $J$-band SED as well as the strong water vapor absorption at 
$\phi \simeq$ 0.73 are indicative of subtraction of too cool a secondary star 
at those phases. In Fig. 8b, we display the phase-resolved spectra from the 
2005 extreme low-state. Like the 2006 low-state the SE spectrum was found by 
averaging three dim-phase spectra together. While weak cyclotron emission was 
observed during the 2006 low-state none was seen during the 2005 extreme 
low-state. 
  
Previous work has determined that ST LMi is a one pole accretor with 
55$^{\circ}$ $\leq i \leq$ 64$^{\circ}$ and 140$^{\circ}$ $\leq \beta \leq$ 
150$^{\circ}$ (Schmidt et al., 1983; Potter, 2000). The primary accretion 
region also appears to have some structure. FBW93 computed 
cyclotron models for ST LMi in a high state, finding that two separate 
emission regions were needed to adequately model the observed spectra from 
the primary pole: the first is a high temperature/high density region located 
between magnetic longitudes ($\psi$) 130 and 170 that has kT = 12.0 keV, and 
log$\Lambda$ = 7.6. 
The second is located between 170 $\leq \psi \leq$ 250 with 
kT = 5 keV and log$\Lambda$ = 4.4. Both regions had magnetic field strengths of
$\simeq$ 12.0 MG, a result which is consistent with the values previously 
published. The relevancy of these models to our SPEX data is unclear since they
were determined when the object was in a high state ($J$ = 13.8) more 
similar to that observed in our SQIID photometry than our low-state 
spectroscopy. In addition, Peacock et al. (1992) reported seeing a second 
pole in ST LMi during a high-state ($B$ $\sim$ 16.7). Their $H$-band 
photometry showed a sudden increase of $\Delta$$H$ = 0.3 mag at 
$\phi$ = 0.35 was observed, with the $J$-band showing a smaller increase. 
Simultaneous polarization curves were published along with their photometry 
that found V/I = 15 - 20\% 
over the duration of the alleged secondary pole. The errors, however, were 
extremely large and the data are in fact, consistent with zero polarization. 
Because both the phasing and amplitude of the 
lightcurve variations are consistent with the WD2005 models found in this 
study (Fig. 5), we believe the excess $H$-band feature may be due to 
ellipsoidal variability from the secondary star.

We used the published values $i$ and $\beta$ to constrain the orbital variation
of $\Theta$ and thus, effectively limit
the possible parameter space to three dimensions, B, kT and log$\Lambda$. 
Because of the very low rate of accretion, the models presented here have low
temperatures. We found 12.0 $\leq$ B(MG) $\leq$ 12.2, 3.2 $\leq$ kT(keV) 
$\leq$ 3.4, and 5.5 $\leq$ log$\Lambda$ $\leq$ 6.1, and $\Theta$ which is well 
described with $i$ = 55$^{\circ}$, $\beta$ = 128$^{\circ}$, $\phi_{max}$ = 
0.22$\pm$ 0.05. The average $\chi^{2}_{\nu}$ = 2.70. As for AM Her , we find 
the following uncertainties on the parameter values: B $\pm$ 0.5 MG, kT $\pm$ 
1.8 keV, and log$\Lambda$ $\pm$ 0.6. Table 3 lists the phase-resolved 
parameters.

\section{Discussion}

\subsection{Analysis of Results}

We have presented the first phase-resolved IR spectroscopy published for 
these two sources allowing for enhanced leverage over the parameters modeled. 
The data presented in this paper proved the most challenging to model in this
series. In Paper I, we covered the basics of cyclotron modeling and applied our
technique to NIR spectra of EF Eri finding that it could be modeled using 
only cyclotron and WD emission. In Paper II, we found that other sources of 
contamination could be eliminated by subtracting the dim phase spectrum from 
each phase for which cyclotron emission was observed. In the present work, 
however, the spectra of both AM Her and ST LMi are dominated by emission from 
the secondary star in the NIR. Because of the orbitally modulated nature of 
both the spectral type and brightness of the secondary, additional issues 
arose. Naive subtraction of only a single dim-phase spectrum from every other 
phase is manifested in two ways. First, variable water vapor 
absorption/emission is seen at at 1.35, 1.85 $\mu$m. Second, the underlying SED
cycles twice between a red and blue excess over the orbit. Both artifacts 
result from the changing spectral type of the secondary star. Because we 
subtract the same phase from each spectrum, the secondary imprint in the 
SE-spectrum alternates between being too cool and too hot when compared to the 
features seen at other phases. To assuage the situation, we median combined 
three dim-phase spectra separated by $\Delta \phi$ = 0.25, to produce the final
SE spectrum, thus smearing out the effects of a changing secondary star. 

In AM Her, we found that for the bright-phase of the 2006 low-state 
(V $\simeq$ 15.5) SPEX data: B = 13.7 $\pm$ 1.0 MG, kT = 4.2$\pm$ 1.0 keV, and 
log$\Lambda$ = 5.0$\pm$ 0.5. The result is in dramatic contrast with
that found in BFW91 at a time when the system was also in a low, though 
perhaps slightly higher state (V $\simeq$ 15.0): B = 14.5 $\pm$ 
0.3 MG, kT = 8.5 $\pm$ 0.5 keV, and log$\Lambda$ = 3.3 $\pm$ 0.3.  While the 
magnetic field strengths for the two epochs agree to within their errors, the 
same cannot be said for both the temperatures and values of log$\Lambda$. Our 
SPEX data show a system with a cooler plasma than that inferred by BFW91. To 
quantify this difference, we evoke the results of  Fischer \& Beuermann 
(2001), which found kT$_{max}$ $\propto$ $\dot{m}B^{-2.6}$, which we rewrite 
as: $\dot{m_{1}}/\dot{m_{2}}$ = ($kT_{1}/kT_{2}$) $(B_{1}/B_{2})^{2.6}$. 
Plugging in the average bright-phase parameters from the two epochs yields that
the $\dot{m}$ for the modeled SPEX data must be a factor of 2.34 lower than 
that active during the BFW91 observations. The higher temperatures found in
BFW91 may be an artifact of medianing together moving cyclotron harmonics over
$\sim$ 40 \% of the orbit, which artifically broadens each feature.

The final spectra and models allow us to understand the changing morphology 
of AM Her's NIR light-curves. In Fig. 3b, the $J$-band is well
explained by ellipsoidal variations alone, while in the $H$ + $K$-bands the 
cyclotron emission component is substantial, with a maximum contribution of 
0.25 mag in both bands and disappearing at $\phi$ = 0.00. The 
$K$-band cyclotron component is relatively constant over the phases 
0.75 $\leq \phi \leq$ 0.25. Such behavior is explained by the cyclotron models
shown in Fig. 4. Near $\phi$ = 0.00 the $n$ = 4 harmonic dominates due to 
the low viewing angle ($\Theta \simeq 35^{\circ}$) at that time. Consequently, 
no emission is seen in the $H$-band. Later, (0.25 $\leq \phi \leq$ 0.75) the 
viewing angle is larger causing the higher harmonics ($n$ = 5 and 6) to be 
excited and thus a peak in the $H$-band emission is observed. Because the $n$ =
4 harmonic is mostly optically thick, however, the cyclotron emission in the 
$K$-band remains relatively constant over the entire bright-phase. 

The simultaneous optical photometry are also interesting. In the $R$ and $I$ 
bands, the data were entirely explained by the ellipsoidal models alone for 
the first full orbit of phase coverage, in line with expectations from our 
13.7 MG cyclotron models that have very little emission shortward of 
1.2 $\mu$m. Subsequently, a brightening event seems to have occurred with 
$\Delta$m = 0.20 in both bands. The $B$ and $V$-bands are more complex and 
cannot be explained by ellipsoidal models alone. Additional modulation was 
observed at the 
level of $\Delta$m = 0.22 and 0.08 for $B$ and $V$, respectively. Gansicke et 
al. (1998) modeled similar UV lightcurves as a hot-spot. In that work, three UV
bands covering wavelength regions of 1150 - 1167$\AA$, 1254 - 1286$\AA$, and 
1412 - 1427$\AA$ were found to be consistent with a 47000 K hotspot centered on
$\psi$ = 0$^{\circ}$ and covering 9 \% of the WD surface. We find the
similarity of our $B$-band lightcurve to Gansicke et al.'s UV lightcurves to 
be striking. Both lightcurves show identical phasing and have amplitudes that
are consistent. We note, however, the large value of $B$ - $V$ $\simeq$ 0.15 
during the bright phase.  For any reasonable hotspot, $B$ - $V$ should be 
closer to 0.00. Because (a) the UV lightcurves predict a very similar geometry 
($i$, $\beta$, and phase) to our cyclotron emission region and (b) the limited
wavelength coverage of the UV lightcurves, we find that this excess emission 
could be caused by a partially saturated high-field cyclotron harmonic ($n$ = 3
or 4) that falls off toward the blue end of the $V$-band and extends through 
the bluest UV band. Such a broad harmonic ($\Delta \lambda$ = 0.4 $\mu$m) is 
expected in cyclotron emission (see AM Her's $n$=4 harmonic in Fig. 4, which is
more or less flat from 1.9 $\mu$m to 2.4 $\mu$m.) If the emission were from the
$n$ = 4 harmonic it would imply a $\sim$ 90 MG field. We also speculated on the
presence of a similar secondary high-field pole for EF Eri in Paper I.

During the bright phase of the 2006 low-state, ST LMi displayed cyclotron
emission with the following properties: B = 12.1 $\pm$ 0.5 M, kT = 3.3 $\pm$ 
1.8 keV and log$\Lambda$ = 5.7 $\pm$ 0.6 similar to the ``cool spot'' found 
in FBW93. In addition, our accretion region appears to be in a similar 
location on the WD surface: at magnetic longitude $\psi$ = 120$^{\circ}$, 
lagging behind the onset of the secondary accretion region found in FBW93 by 
$\Delta\phi$ $\simeq$ 0.13, likely due to the accumulation in phase-error 
between 1991 and 2005. Unfortunately, no errors are given for any of the
cyclotron parameters in FBW93 and thus the significance of the difference in 
results is hard to assess. However, both the FBW93 magnetic field strength, and
the plasma temperature agree to within our errors. More interesting is the 
non-detection of their primary accretion region which should trail the 
observed ``secondary'' emission region by about 0.10 in phase, implying an 
onset at $\phi$ $\simeq$ 0.60 which is not seen. 

\subsection{Ellipsoidal Versus Cyclotron Derived Inclinations}

In both cases, the geometry of the emission region was consistent with a 
simple single spot model having a constant orbital inclination and magnetic 
colatitude. In Fig. 9, we plot (black) the cyclotron derived values of the 
viewing angle against the orbital phase for both AM Her(top) and 
ST LMi(bottom). In red are the simple geometrical models, with the blue shading
indicating phases for which the cyclotron regions are self eclipsed. The 
models fit the AM Her data well for nearly all phases, only deviating 
near self-eclipse ingress and egress, where the viewing angle is 
changing rapidly compared to the cadence of our spectroscopy. For ST LMi, the 
bright phase is relatively short, lasting $\simeq$ 40 \% of the orbit. 
Consequently, few data were available to constrain its geometry. We thus used 
published values of $i$ and $\beta$ for an additional constraint, finding that 
we could match the data with models quite similar to those found in the 
literature.

For AM Her, the cyclotron models imply an orbital inclination of 
$i$ = 50$^{\circ}$, identical to that found in the 
ellipsoidal modeling effort, although higher angles are possible if additional
sources of dilution remain in the IR light curves. Agreement was also found for
the ellipsoidal and cyclotron inclinations for ST LMi finding values of $i$ = 
55$^{\circ}$ and 40$^{\circ}$, respectively. The later value represents the 
lowest inclination ellipsoidal model, and values of $i$ = 55$^{\circ}$ are more
consistent with the lightcurve (see Fig. 3). Some caution, however, should be 
given to the fact that the spectroscopy and lightcurves of ST LMi were taken 
at different epochs when the object was in different states, which could
affect the accretion geometry. In Schwope et al. (1993), the authors 
found that the polar MR Serpentis appeared to show longitudinal migration of
the accretion spot by $\simeq$ 30$^{\circ}$, as well as a 10$^{\circ}$ shift
in the magnetic colatitude between its high and low states.       

\subsection{ST LMi in an Extreme Low-State}

In Fig. 8b, we present SE-Subtracted SPEX data obtained during our 2005 
February 7 observing run, which show a conspicuous lack of cyclotron emission
throughout the entire orbital cycle. The extreme low-state of ST LMi is 
corroborated by near-epoch (2006 February 12) optical lightcurves obtained 
with the WIYN 0.9-m, showing the system in a deep low-state. Despite the poor 
telluric correction (the spectra were faint), the only strong feature in the 
bright-phase (0.50 $\leq \phi \leq$ 0.85) is small ``bump'' longward of 
2.2 $\mu$m, caused by under subtraction of the secondary star at those phases. 
In this series of papers (see Papers I,II, and Szkody et al. 2008, submitted) 
we have modeled seven polars, representing $\sim$ 10\% of all known mCVs. 
Included in this sample were EQ Cet, the prototype ``Low-Accretion Rate 
Polar'', MQ Dra (= SDSS 1553), the prototype ``pre-polar'', and EF Eri, well 
known for its protracted low-state. Intriguingly, in each of these objects 
strong cyclotron emission was observed, while the extreme low-state dataset of 
ST LMi is the only example where cyclotron emission completely disappeared. A 
similar situation probably explains the 2006 Feb 12 photometric low-state found
by Kafka et al. (2007). This suggests that normal polars can have periods that 
appear completely devoid of detectable accretion.

\section{Conclusion}
We obtained a full orbit of phase-resolved IR spectra for both AM Her and 
ST LMi. We found both objects to be dominated by emission from the secondary
in the IR. To remove this component, we utilized the fact that emission regions
for both stars were self-eclipsed. Thus, at each phase we subtract a dim-phase
or ``Stream-Emission'' spectrum. Because of the changing spectral type of
the secondary, we found that medianing dim-phase spectra over $\simeq$ 25 \%
of an orbit produced a better subtraction. For AM Her, we found a phase 
averaged model of :  B = 13.6$_{-0.8}^{+1.0}$ MG, kT = 4.0$_{-1.0}^{+1.5}$ 
keV (see Table 2 for specifics at each phase). Additionally, we found that the 
viewing angle varied in a manner consistent with expectations from a system 
with $i$ = 50$^{\circ}$ and $\beta$ = 85$^{\circ}$. For ST LMi, we collected 
two datasets. The first had ST LMi in a low-state with $V$ $\simeq$ 17.4, and
displayed weak cyclotron harmonics that were difficult to decouple from
the water vapor signatures leftover after SE-subtraction. We found a phase 
averaged model of :  B= 12.1 $\pm$ 0.5 MG, kT = 3.3 $\pm$ 1.8 keV in an 
accretion region consistent with $i$ = 55$^{\circ}$ and $\beta$ = 
128$^{\circ}$. For ST LMi, we include a second data set, taken when the object 
was in an ``extreme low-state'' showing no substantial cyclotron 
emission. The non-detection of cyclotron emission contrasts with our earlier 
data from both EQ Cet and MQ Dra both of which show cyclotron emission, even 
while being in extremely low-states. 

\acknowledgments

\clearpage
\begin{deluxetable}{llllll}
\tabletypesize{\scriptsize}
\tablecaption{Observing Log}
\tablewidth{0pt}
\tablehead{\colhead{Date} & \colhead{Object}  & \colhead{Instrument} 
&\colhead{Obs. Type} &\colhead{I(sec)} &\colhead{State}}
\startdata

2005 September 1  &  AM Her & IRTF/SPEX & Spec. & 120 & Low \\
2002 September 26 &  AM Her & KPNO      & Phot. & & Low \\
2005 May 20       &  AM Her & APO       & Phot. & 240 & Low \\
2005 February 7   &  ST LMi & IRTF/SPEX & Spec. & & Extreme Low \\
2006 February 2   &  ST LMi & IRTF/SPEX & Spec. & & Low \\
2003 April 9      &  ST LMi & KPNO      & Phot. & 240 & High \\

\enddata

\end{deluxetable}
\clearpage

\begin{deluxetable}{llllll}
\tabletypesize{\scriptsize}
\tablecaption{Cyclotron Modeling Parameters for AM Her}
\tablewidth{0pt}
\tablehead{\colhead{Phase} & \colhead{B(MG)} &\colhead{kT(keV)}
&\colhead{$\theta$} &\colhead{log$\Lambda$} &\colhead{$\chi^{2}_{\nu}$} }
\startdata

0.03 &  13.3 & 3.9 & 35.0 & 5.3 & 3.53 \\
0.09 &  13.1 & 4.1 & 42.0 & 5.3 & 3.90 \\ 
0.20 &  14.0 & 3.9 & 62.0 & 4.8 & 1.11 \\
0.26 &  14.1 & 3.9 & 72.0 & 4.6 & 2.12 \\
0.31 &  $\cdots$ &  $\cdots$ &  $\cdots$ &  $\cdots$ & $\cdots$ \\
0.37 &  $\cdots$ &  $\cdots$ &  $\cdots$ &  $\cdots$ & $\cdots$ \\
0.48 &  $\cdots$ &  $\cdots$ &  $\cdots$ &  $\cdots$ & $\cdots$ \\
0.52 &  $\cdots$ &  $\cdots$ &  $\cdots$ &  $\cdots$ & $\cdots$ \\
0.59 &  $\cdots$ &  $\cdots$ &  $\cdots$ &  $\cdots$ & $\cdots$ \\
0.64 &  $\cdots$ &  $\cdots$ &  $\cdots$ &  $\cdots$ & $\cdots$ \\
0.70 &  $\cdots$ &  $\cdots$ &  $\cdots$ &  $\cdots$ & $\cdots$ \\
0.75 &  13.8 & 4.3 & 75.0 & 5.0 & 1.78 \\
0.80 &  13.6 & 4.3 & 67.0 & 5.2 & 1.30 \\
0.86 &  13.6 & 4.2 & 55.0 & 5.1 & 3.30 \\
0.92 &  13.4 & 4.0 & 46.0 & 5.1 & 2.28 \\
0.97 &  13.3 & 3.9 & 37.0 & 5.4 & 2.48 \\

\enddata

\end{deluxetable}
\clearpage

\begin{deluxetable}{llllll}
\tabletypesize{\scriptsize}
\tablecaption{Cyclotron Modeling Parameters for ST LMi}
\tablewidth{0pt}
\tablehead{\colhead{Phase} & \colhead{B(MG)} &\colhead{T(keV)}
&\colhead{$\theta$} &\colhead{log$\Lambda$} &\colhead{$\chi^{2}_{\nu}$} }
\startdata

0.14 &  $\cdots$ &  $\cdots$ &  $\cdots$ &  $\cdots$ & $\cdots$\\
0.26 &  $\cdots$ &  $\cdots$ &  $\cdots$ &  $\cdots$ & $\cdots$ \\
0.38 &  $\cdots$ &  $\cdots$ &  $\cdots$ &  $\cdots$ & $\cdots$\\
0.50 &  $\cdots$ &  $\cdots$ &  $\cdots$ &  $\cdots$ & $\cdots$\\
0.57 & 12.0 & 3.2 & 85.0 & 5.5 & 2.19 \\
0.68 & 12.0 & 3.2 & 78.0 & 5.5 & 2.61 \\
0.73 & 12.2 & 3.4 & 75.0 & 6.1 & 3.75 \\
0.85 & 12.2 & 3.4 & 82.0 & 5.9 & 2.24 \\
0.91 &  $\cdots$ &  $\cdots$ &  $\cdots$ &  $\cdots$ & $\cdots$\\
0.99 &  $\cdots$ &  $\cdots$ &  $\cdots$ &  $\cdots$ & $\cdots$\\

\enddata

\end{deluxetable}
\clearpage

{\bf Figure 1.}
$JHK$ photometry of AM Her obtained with the KPNO 2.1-m on September 26, 2002 
when the object was in an faint-state ($V$ $\sim$ 15.5). The $J$ and 
$H$ bands show strong ellipsoidal variations, while the $K$-band morphology
is the result of a combination of ellipsoidal and cyclotron emission. The 
lightcurves were phased using the Kafka et al. (2005b) ephemeris. The 
overplotted lines are ellipsoidal models for $i$ = 50$^{\circ}$.

{\bf Figure 2.} Long-term RoboScope lightcurves of AM Her complimented by 
AAVSO data. The ``P'' denotes the times of our our SQIID and NIC-FPS/NMSU 1-m 
photometry, while the `` S'' indicates our spectroscopic observations. top: 
$V$-band photometry following AM Her from 1991 through 2006. bottom-left: 
Zoom-in of the year surrounding our SQIID photometric measurements. 
bottom-right: Zoom-in of the year surrounding our IRTF spectroscopy. 

{\bf Figure 3.}
$BVRIJHK$ photometry of AM Her obtained with the APO 3.5-m/NMSU 1-m on May 20, 
2005 when the object was in a faint-state (V = 15.3) similar to the 
KPNO lightcurves (see Fig. 1). An identical ellipsoidal model to that used for 
the KPNO photometry is overplotted here, matching well from $R$ to $K$, and 
although a small flare event is evident in $R$ and $I$ during the second cycle 
of observation almost no cyclotron emission should be present in these bands. 
Humps reminiscent of cyclotron emission reappear in the $V$ and especially $B$ 
bands suggesting that a higher field is active on AM Her. (a) The optical
bands (b) the NIR bands

{\bf Figure 4.}
(a)IRTF/SPEX phase-resolved spectra of AM Her plotted (black) as
a stacked series, with a constant flux increment of  $\lambda F_{\lambda}$ = 
1.10$\times$10$^{-11}$ erg s$^{-1}$ cm$^{-2}$ and covering the
orbital phases for which the cyclotron emission region is in view. At each 
phase, the 1.22 $\mu$m flux is normalized to the $J$-band lightcurve ensuring 
proper flux calibration with the narrow 0.3'' slit and a dim-phase 
spectrum ($\phi$ = 0.42) was subtracted.  Because of the variability of the 
secondary's spectral type over the orbit, the underlying continuum as well as 
intrinsic water features changed over the orbit. Remnant intrinsic water 
features, however, are still apparent at $\sim$ 1.35 and 1.8 $\mu$m. No 
cyclotron emission was observed from $\phi$ = 0.27 to $\phi$ = 0.74.

{\bf Figure 5.}
$JHK$ photometry of ST LMi obtained with the KPNO 2.1-m/SQIID on the April 
9, 2003 high-state. The dim-phase lasts from $\phi$ = 0.00 to 0.55, while from
$\phi$ = 0.60 to 0.95 the bright-phase is observed. The plotted lines are 
ellipsoidal models for $i$ = 55$^{\circ}$.

{\bf Figure 6.} Long-term lightcurves of ST LMi. top: $V$-band photometry 
following ST LMi from late 1991 to early 2006. bottom: Zoom-in of the year
surrounding our IRTF/SPEX spectroscopy. The ``P'' denotes the epoch of our 
photometry, while ``S'' marks our spectroscopic data during the 2005 extreme 
low-state and the 2006 low-state, respectively.

{\bf Figure 7.}
IRTF data of ST LMi in an extreme low-state (2005). An M6 template spectrum is 
plotted during the dim-phase ($\phi$ = 0.02) confirming the spectral 
classification.

{\bf Figure 8.}
(a)Phase-Resolved SE-subtracted spectroscopy of ST LMi, taken in February 2006
during a low-state. The IRTF/SPEX data are plotted (black) as
a stacked series - a constant increment of $\lambda F_{\lambda}$ = 
1.2$\times$10$^{-12}$ erg s$^{-1}$ cm$^{-2}$ is added to each spectrum
to offset it from the spectrum below it. The SE spectrum subtracted from
each phase was a median of three dim-phase spectra. The best fit
cyclotron model for each of the bright-phase spectra are shown in green. (b)
the same, but for the February 2005 extreme low-state.

{\bf Figure 9.} (a) The derived value of the viewing angle ($\Theta$) for
AM Her is plotted vs. the orbital phase ($\phi$) in black, with the best fit 
geometry ($i$ = 50$^{\circ}$, $\beta$ = 85$^{\circ}$) overlaid in red. The blue
shading indicates phases for which AM Her is self-eclipsed. (b) the same, but
for ST LMi and with a geometrical model of ($i$ = 55$^{\circ}$, $\beta$ = 
128$^{\circ}$).

\begin{figure}
\epsscale{0.80}
\plotone{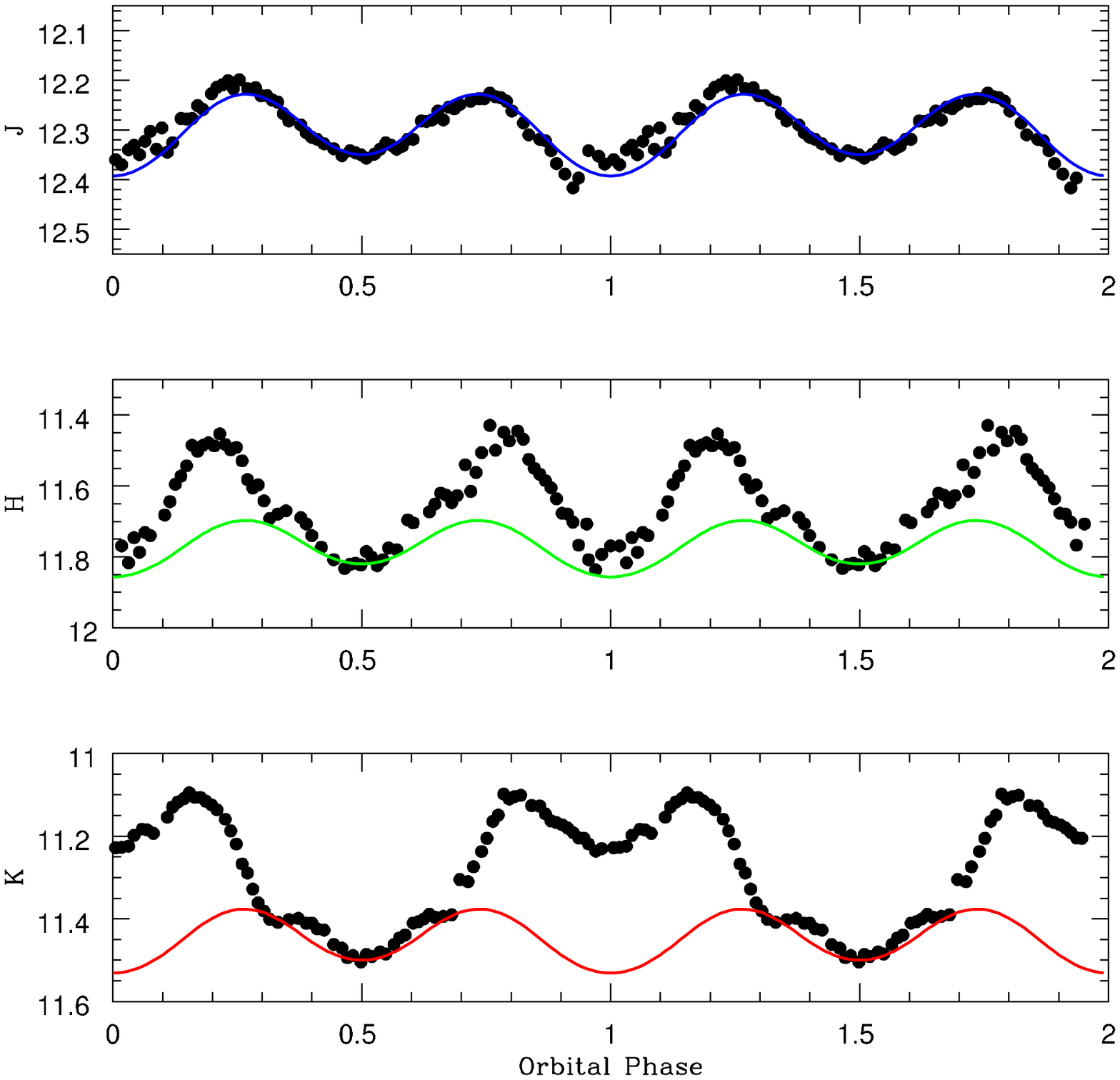}
\end{figure}
\begin{figure} 
\epsscale{0.80}
\plotone{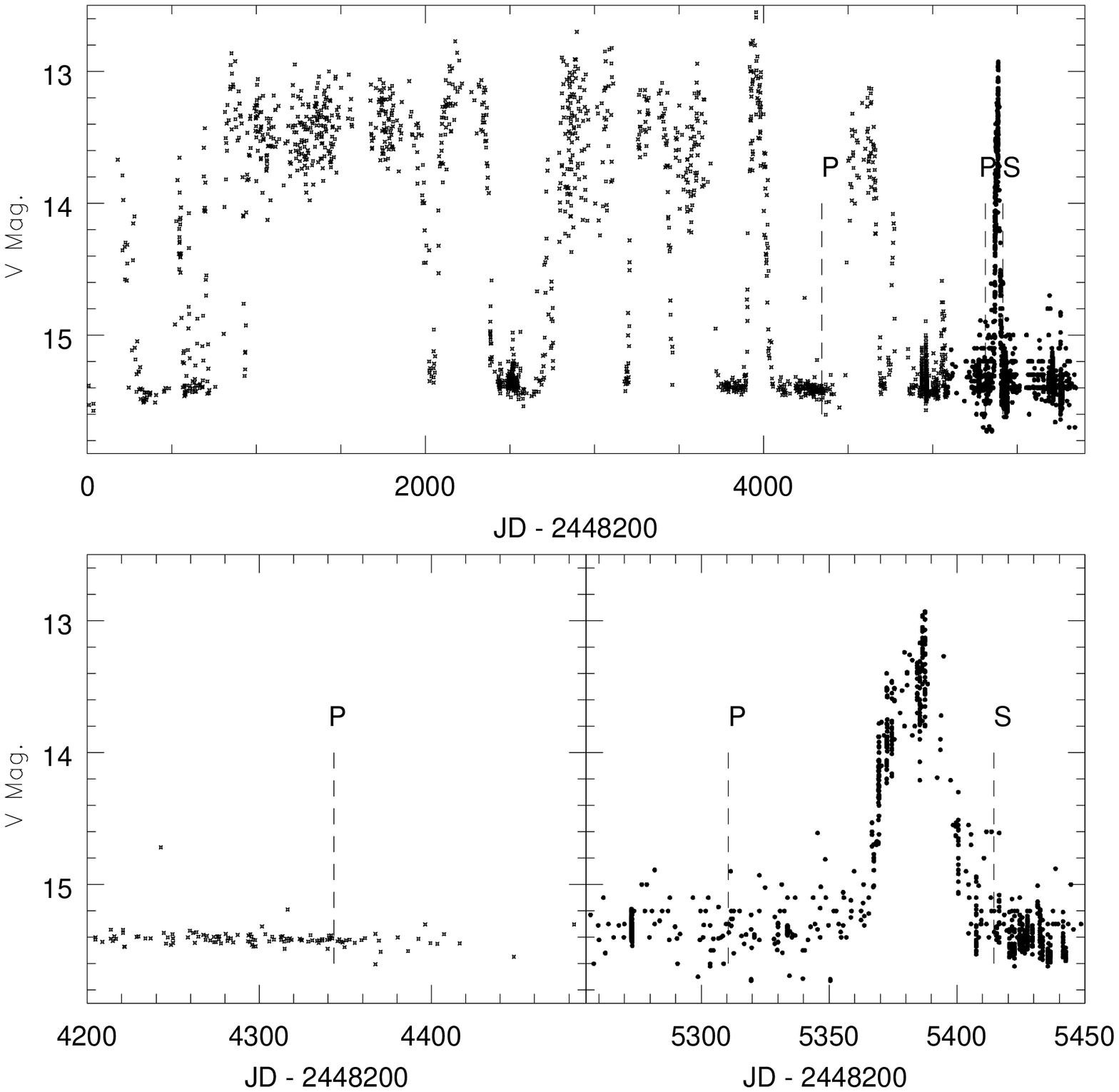}
\end{figure}
\begin{figure}
\epsscale{0.80}
\plotone{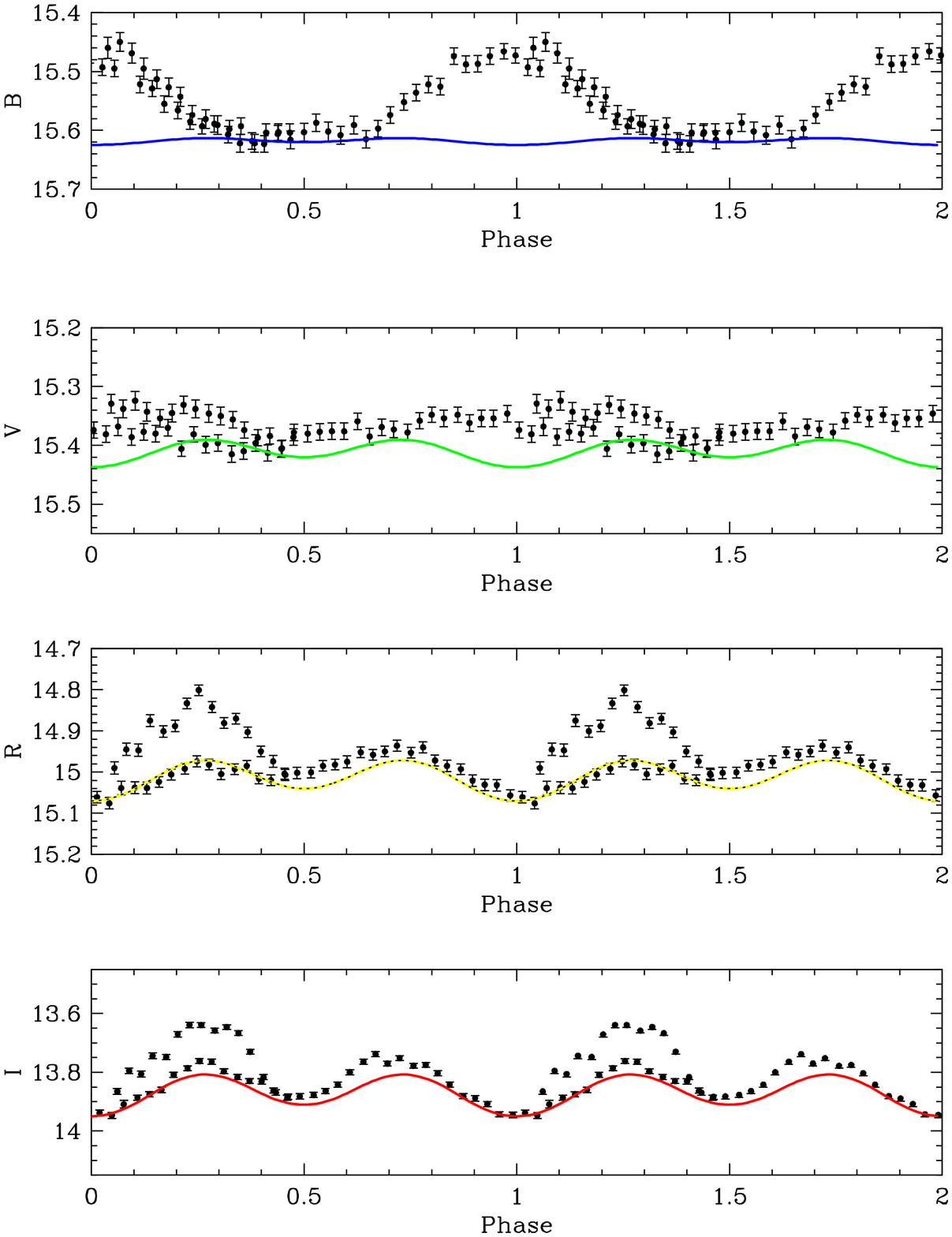}
\end{figure}
\begin{figure}
\epsscale{0.80}
\plotone{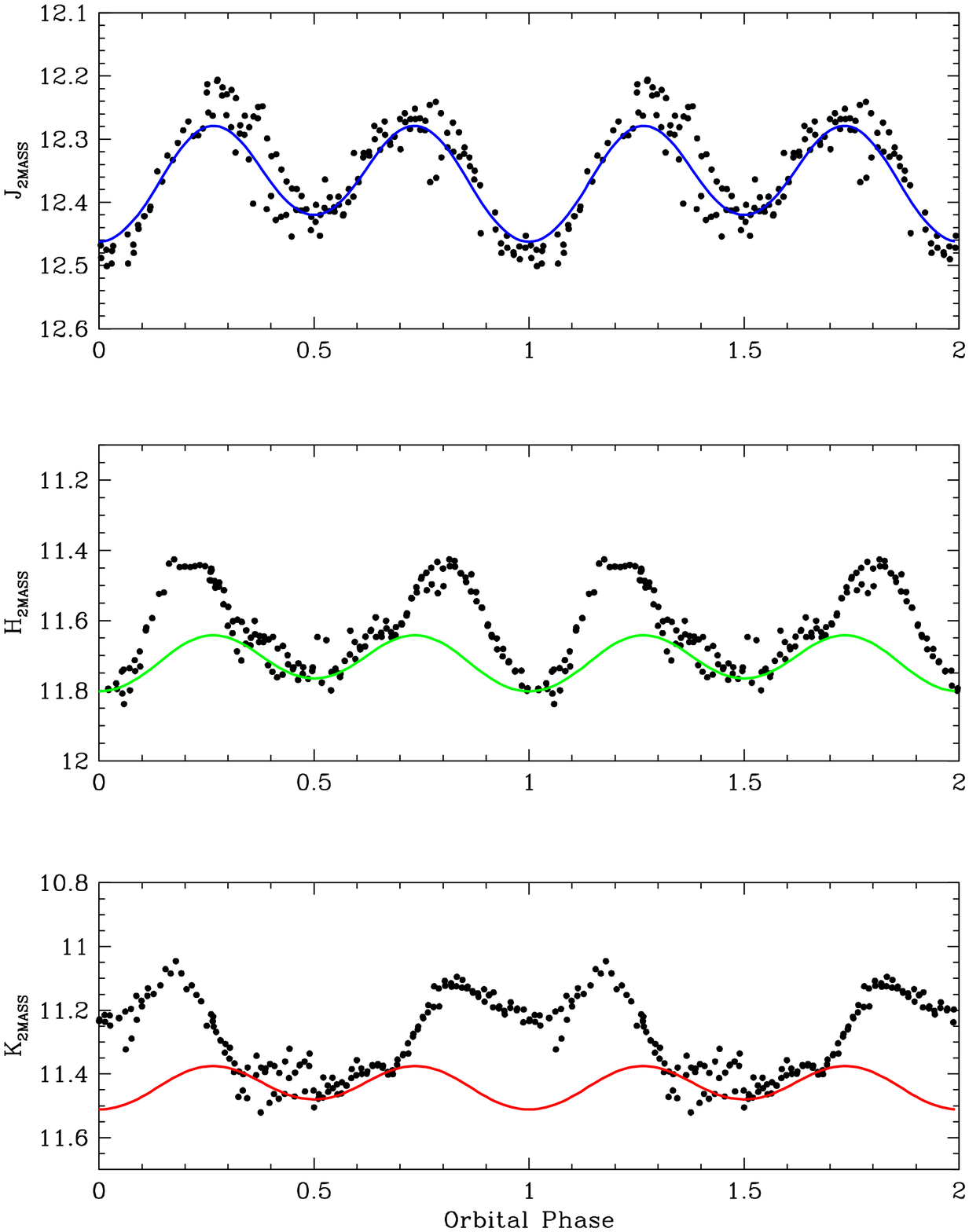}
\end{figure}
\begin{figure}
\epsscale{0.80}
\plotone{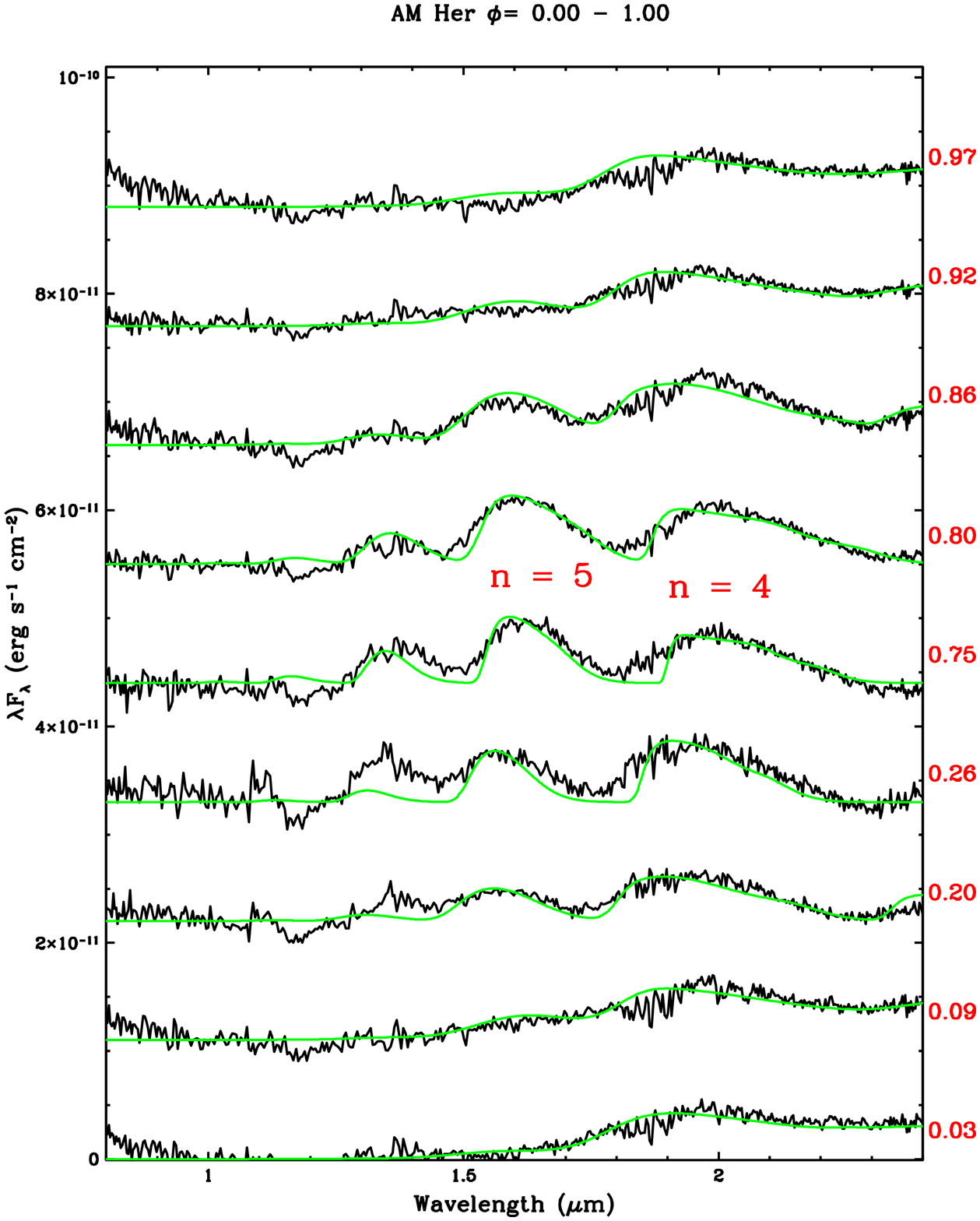}
\end{figure}
\begin{figure}
\epsscale{0.80}
\plotone{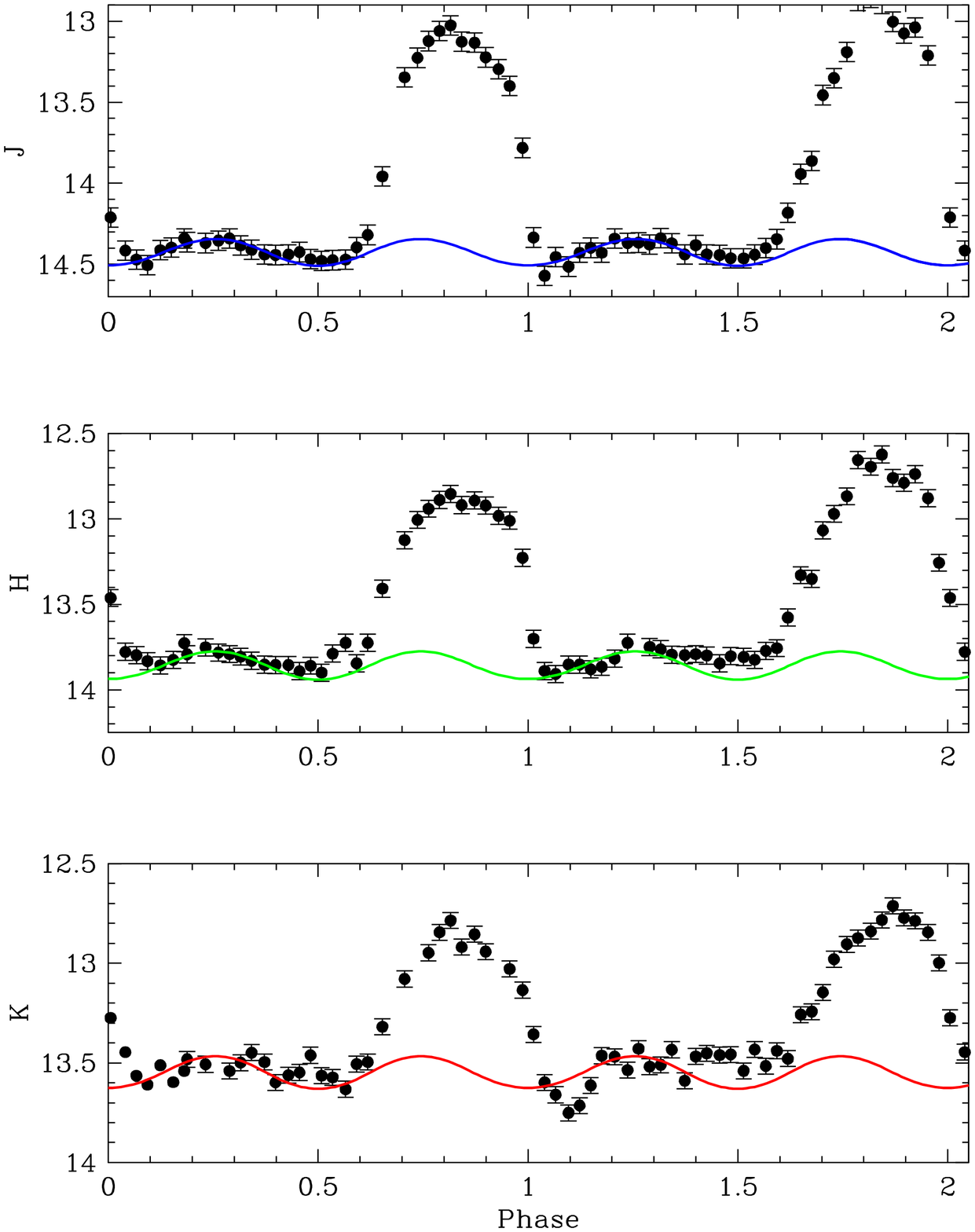}
\end{figure}
\begin{figure}
\epsscale{0.80}
\plotone{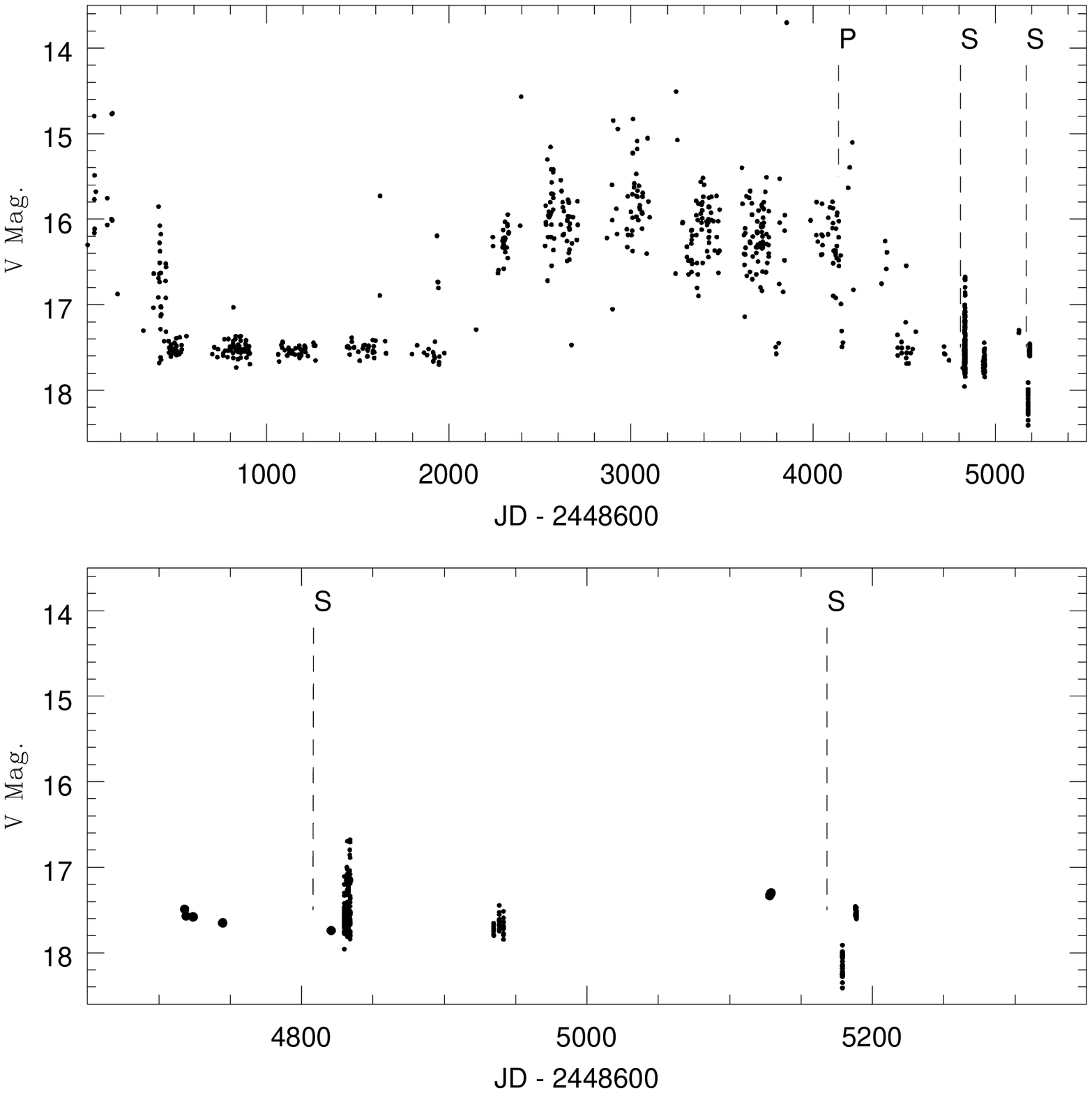}
\end{figure}
\begin{figure}
\epsscale{0.80}
\plotone{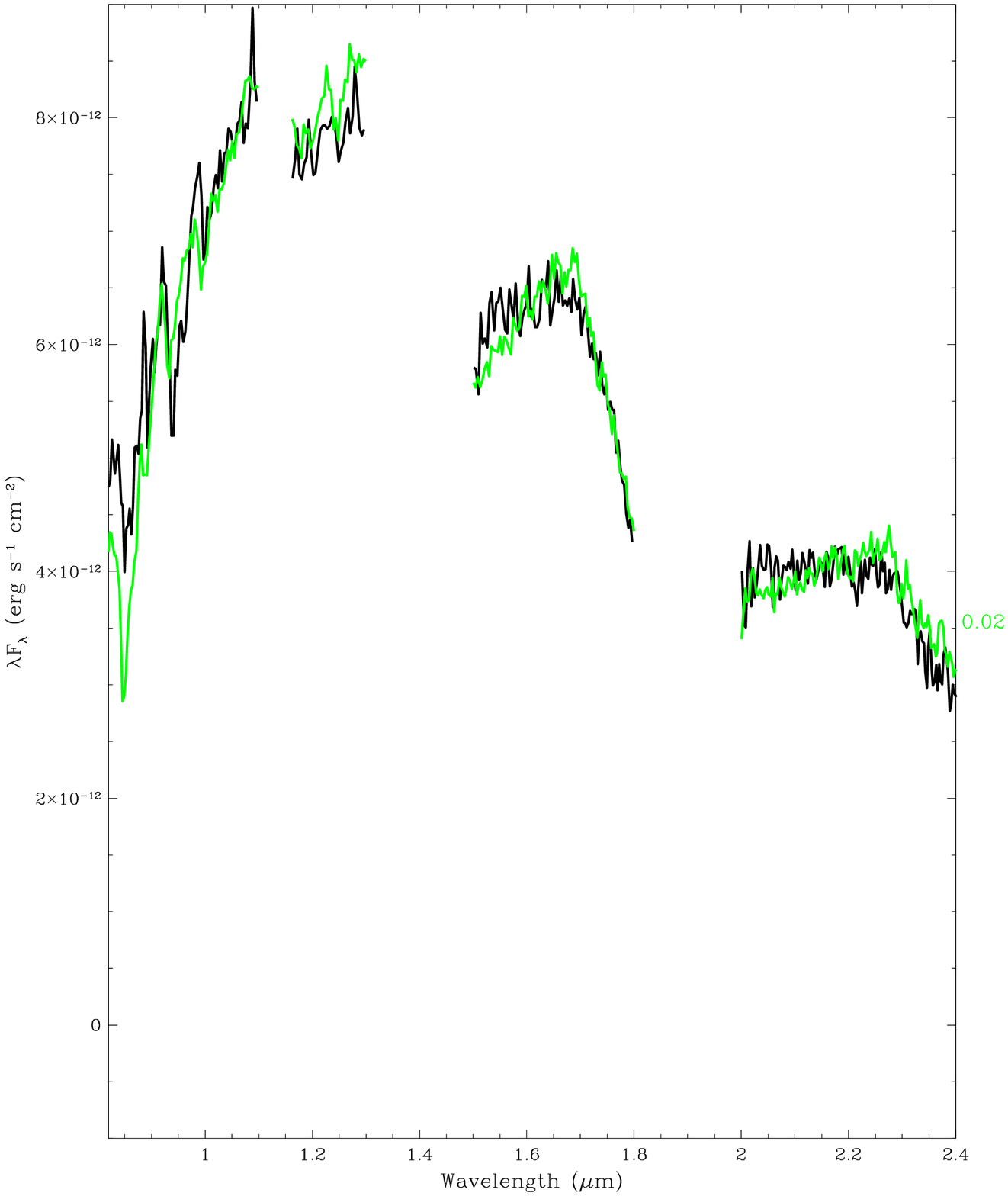}
\end{figure}
\begin{figure}
\epsscale{0.80}
\plotone{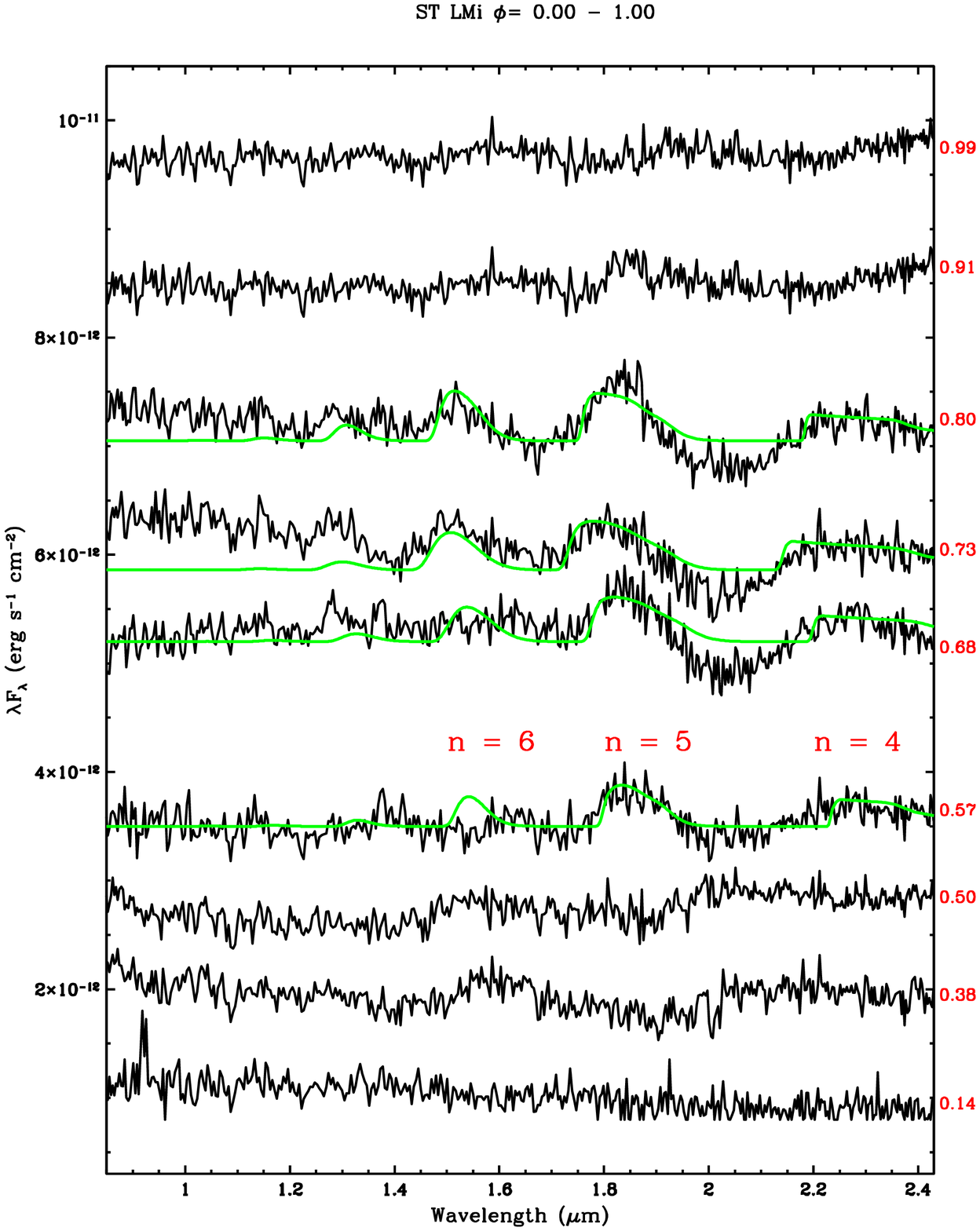}
\end{figure}
\begin{figure}
\epsscale{0.80}
\plotone{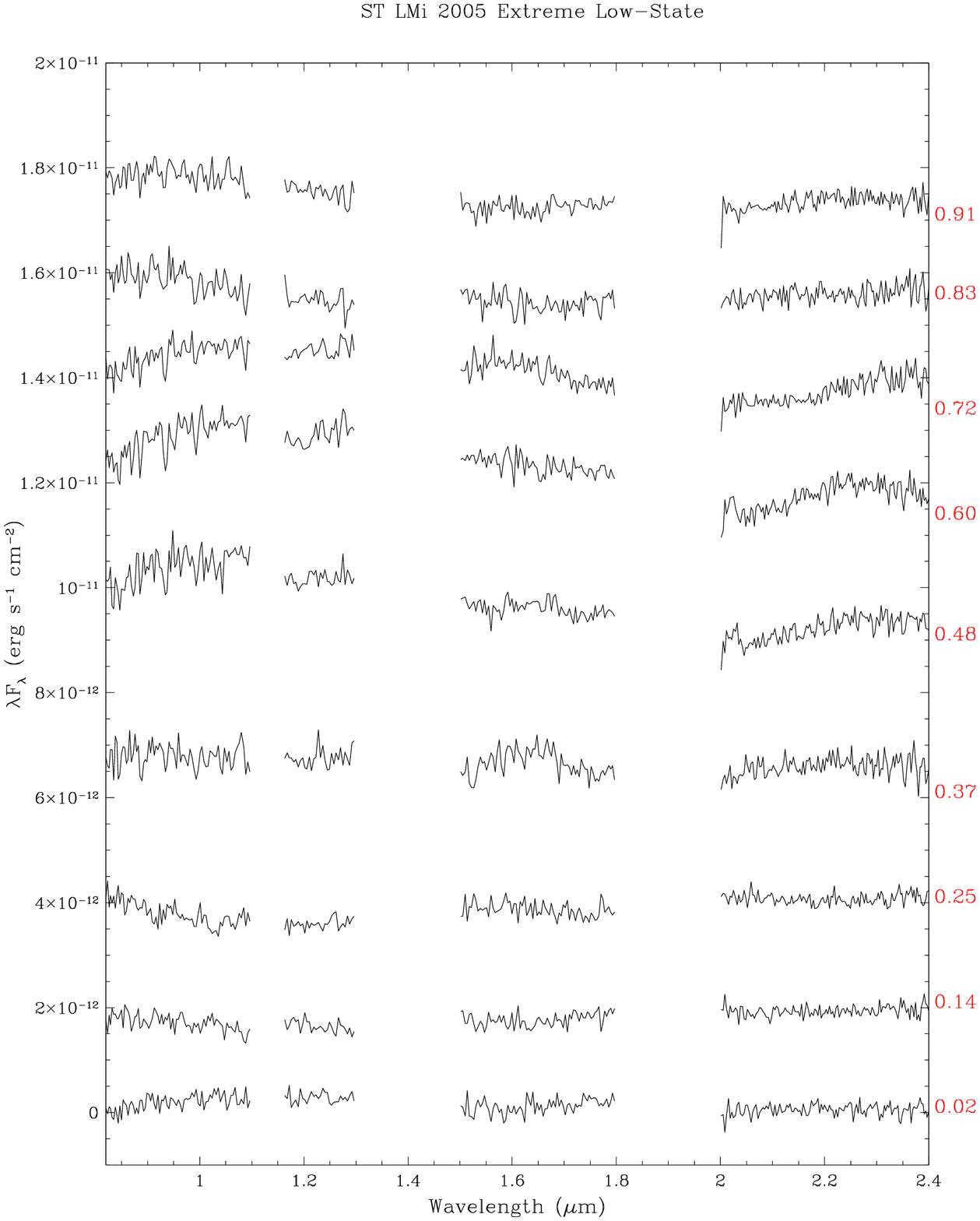}
\end{figure}
\begin{figure}
\epsscale{0.80}
\plotone{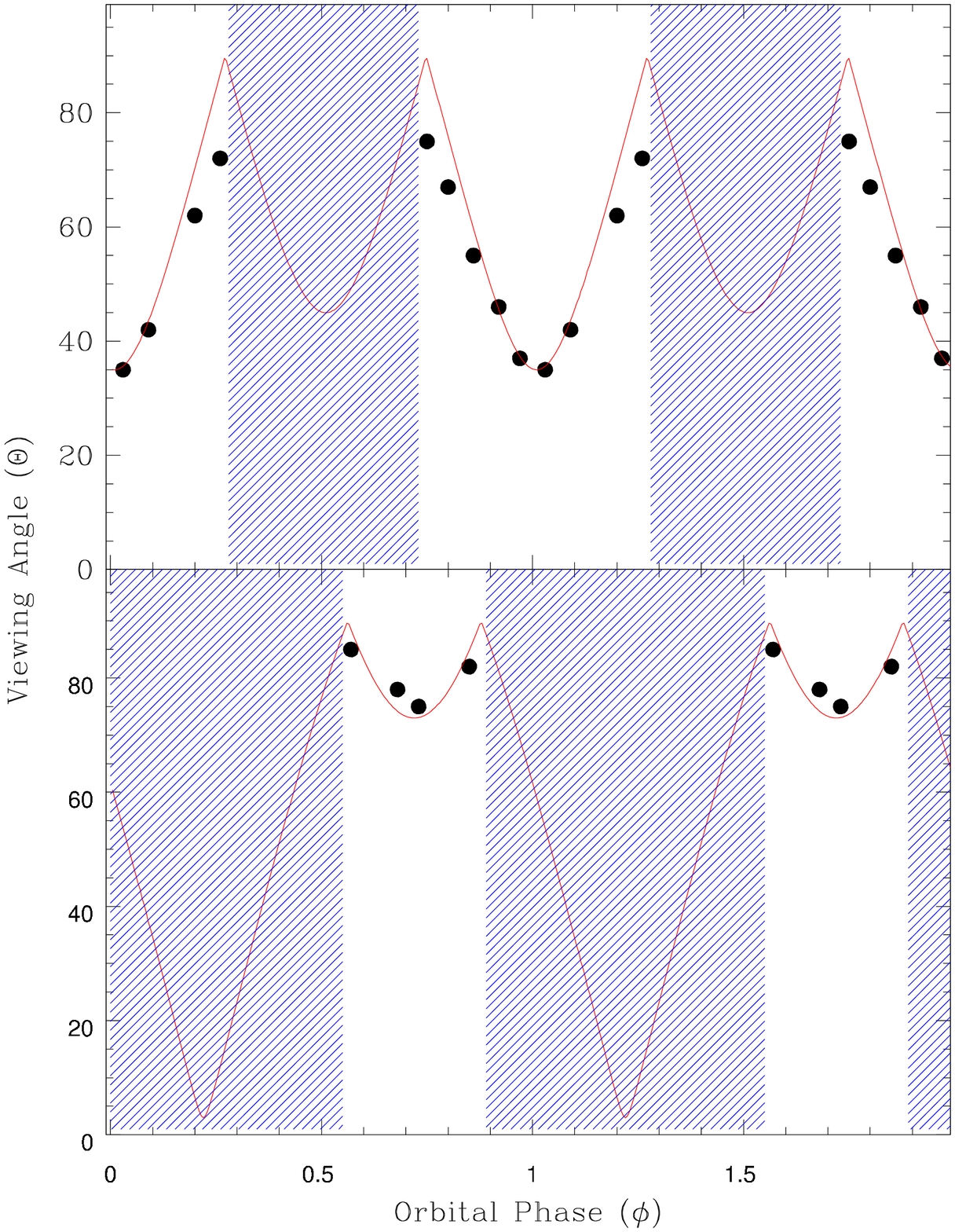}
\end{figure}

\end{document}